\documentclass[12pt,english,floatfix,showkeys,superscriptaddress,aps,prd]{revtex4}
\usepackage[utf8]{inputenc}
\usepackage[english]{babel}
\usepackage[T1]{fontenc}
\usepackage{lmodern}
\usepackage{amsmath}
\usepackage{subfigure}
\usepackage{amssymb}
\usepackage{graphicx}
\usepackage{float}
\usepackage{esint}
\usepackage{longtable}
\usepackage{dcolumn}
\usepackage{babel}
\usepackage{csquotes}
\usepackage{color}
\usepackage{slashed}
\usepackage{simplewick}
\usepackage[utf8]{inputenc}                   
\usepackage{amsmath,latexsym}

\usepackage{hyperref}
\hypersetup{
    colorlinks,
    citecolor=blue,
    filecolor=green,
    linkcolor=purple,
    urlcolor=red,
}

\usepackage{slashed}

\usepackage{hyperref}
\hypersetup{colorlinks,breaklinks,
			citecolor=[rgb]{0,0.0,1.0},
            urlcolor=[rgb]{0.0,0.0,1.0},
            linkcolor=[rgb]{0,0.5,0.9}}

\begin{document}

\title{Hot Casimir wormholes in Einstein Gauss–Bonnet gravity}

\author{C. R. Muniz}
\email{celio.muniz@uece.br}
\affiliation{Universidade Estadual do Cear\'a (UECE), Faculdade de Educa\c{c}\~ao, Ci\^encias e Letras de Iguatu, Av. D\'ario Rabelo s/n, Iguatu - CE, 63.500-00 - Brazil.}


\author{M. B. Cruz}
\email{messiasdebritocruz@servidor.uepb.edu.br}
\affiliation{Universidade Estadual da Para\'iba (UEPB), \\ Centro de Ci\^encias Exatas e Sociais Aplicadas (CCEA), \\ R. Alfredo Lustosa Cabral, s/n, Salgadinho, Patos - PB, 58706-550 - Brazil.}


\author{R. M. P. Neves}
\email{raissa.pimentel@uece.br}
\affiliation{Universidade Estadual do Cear\'a (UECE), Faculdade de Educa\c{c}\~ao, Ci\^encias e Letras de Iguatu, Av. D\'ario Rabelo s/n, Iguatu - CE, 63.500-00 - Brazil.}


\author{Mushayydha Farooq}
\email{Mushayydha_90@hotmail.com}
\affiliation{Department of Mathematics, COMSATS University Islamabad, Lahore Campus, Lahore, Pakistan}


\author{M. Zubair}
\email{mzubairkk@gmail.com; drmzubair@cuilahore.edu.pk}
\affiliation{Department of Mathematics, COMSATS University Islamabad, Lahore Campus, Lahore, Pakistan}
\affiliation{National Astronomical Observatories, Chinese Academy of Sciences, Beijing 100101, China}


\date{\today}

\begin{abstract}
In this work, we explore the thermal effects on Casimir wormholes in the context of higher-dimensional Einstein-Gauss-Bonnet gravity. Motivated by the fundamental role of EGB gravity in describing a wide range of gravitational phenomena, we investigate how thermal fluctuations affect the quantum vacuum energy density associated with the Casimir effect and its impact on the global structure of traversable wormholes. By deriving the shape function from the EGB field equations with thermally corrected Casimir energy, we verify that all necessary conditions for wormhole formation are satisfied, including asymptotic flatness and throat stability. Our results indicate that thermal corrections modify the wormhole’s geometry, increasing spatial curvature in the throat region and influencing its traversability. Furthermore, we analyze gravitational Casimir effects and discuss their possible role in modified gravity theories. Expanding on the approach of reference \cite{M. Zubair1, Mushayydha, Mushayydha2}, we adopt here the appropriate formulation for Casimir wormholes in Einstein-Gauss-Bonnet gravity, taking into account the Casimir energy density in higher dimensions. This approach allows us to obtain more accurate results compared to the simplified approximation previously used.
\end{abstract}


\maketitle


\section{Introduction}
\label{Introduction}

Wormholes are hypothetical "tunnels" in spacetime that could connect two distant locations \cite{Einstein, M.S.Morris}. Their structures have been extensively studied within Einstein's General Relativity (GR) as well as in various modified gravity theories, each incorporating different physical assumptions and background conditions \cite{T. Harko, G. Clement, K.A. Bronnikov, K. Jusufi, K. Jusufi2, K. Jusufi3, K. Jusufi4, F. Rahaman}. The phenomenon of wormholes has been widely explored in several modified gravity theories, such as $f(R)$, $f(T)$, $f(R,T)$, Brans-Dicke theories, and scalar-tensor theories \cite{F.S.N. Lobo, M. Jamil, C.G. Boehmer, A.G. Agnese, K.K. Nandi, F. He, E. Ebrahimi, M. Zubair, S. Bahamonde}. Despite the recent great success of GR, its fascinating prediction regarding wormholes has yet to be directly confirmed through observations.

There are theoretical constructions of wormholes, known as Casimir wormholes, which are based on the energy densities and pressures associated with the Casimir effect \cite{Garattini:2019ivd, Muniz:2024jzg}. Over the past decade, interest in the Casimir force has grown considerably. This quantum phenomenon, which results from the confinement of vacuum fluctuations of electromagnetic fields between two nearby reflective surfaces, was initially predicted by Hendrik Casimir in 1948 \cite{Casimir:1948dh} and checked \cite{Sparnaay:1958wg, Lamoreaux:1996wh, Harris:2000zz} later, plays a crucial role in various areas of physics, such as quantum field theory \cite{Cruz:2017kfo, Cruz:2018thz, Cruz:2020zkc}, nanotechnology, and condensed matter physics \cite{LouysSanso:2025ijj, Shuai:2025zes, Flachi:2024sff}.

To understand the stability, evolution, and physical viability of Casimir wormholes, their dynamics have been extensively studied in recent years within the framework of various modified gravity theories. These investigations analyze how different gravitational models interact with quantum fluctuations, negative energy densities, and the need for exotic matter to support traversable wormholes. In particular, studies have explored the hypothesis that the negative energy density of the Casimir effect could, in theory, allow for the existence of traversable wormholes \cite{Cruz:2024ihb}. Additionally, research on the impact of weak quantum energy conditions on the stability and traversability of these hypothetical structures has also been conducted \cite{Remo Garattini}. Still in the context of Casimir wormholes, several interesting works have been developed. For example, using parallel plates arranged in a circle around rotating wormholes from Teo and Damour-Solodukhin, a study was conducted in \cite{C. R Muniz}. Additionally, considering the Generalized Uncertainty Principle (GUP) and its corrections, several studies have been carried out, as discussed in \cite{Kimet5, Zinnat Hassan, M. Zubair1}. Regarding the effects of electric charge and corrections to quantum gravity, very interesting studies have also been conducted, especially in the context of quantum gravity \cite{Zinnat Hassan2, Mushayydha, Peter K.F. Kuhfittig, Mushayydha2}.

An important factor to consider in the Casimir effect, and consequently in Casimir wormholes, is the finite temperature corrections arising from the interactions between conductive or dielectric surfaces due to thermal and quantum fluctuations. In this regard, for larger distances between the interacting surfaces, thermal contributions can modify both the amplitude of the force and its qualitative behavior compared to the zero-temperature case \cite{J. Marino, A.O. Sushkov, Bostroom}. Studies on the dynamic behavior of wormholes \cite{Remo Garattini} have also addressed this scenario, incorporating thermal corrections \cite{Remo Garattini2, Phongpichit Channuie}.

Studies indicate the need for a type of exotic matter to enable the formation of wormholes, such as Casimir energy. However, this exotic matter violates the energy conditions \cite{M.S.Morris}. As a result, several studies have been conducted with the goal of reducing the dependence on this exotic matter \cite{Lobo, Visser}. Interesting works have shown that the energy requirements necessary in the throat of a wormhole can be satisfied by higher-dimensional cosmological wormholes or those that incorporate aspects of higher-order curvature invariants \cite{Zangeneh, Harko, Dzhunushaliev, PoncedeLeon}.

Einstein-Gauss-Bonnet gravity is a relevant theory within the class of modified gravity theories, commonly known as Lovelock gravity, developed by David Lovelock \cite{Lovelock}. In the context of EGB, terms of higher-order curvature arise; however, the equations of motion maintain a structure with second-order derivatives in the metric. As a result, Lovelock gravity provides a natural application of general relativity (GR) to higher-dimensional spacetimes, preserving fundamental mathematical characteristics and incorporating curvature corrections. Therefore, this theory contains important features that other higher-order curvature modified gravity theories do not possess. The higher-order curvature terms can be interpreted as a type of gravitational fluid, essential for supporting the geometries of wormholes in modified gravity theories. In this way, it is possible to confine matter passing through the throat of the wormhole, ensuring that all energy conditions are satisfied, which does not occur in conventional models. Due to this possibility, the investigation of higher-dimensional wormhole geometries has been strongly recommended.

This work aims to study the thermal effects on Casimir wormholes within the framework of higher-dimensional EGB gravity. The structure of the paper is organized as follows: Section \ref{Basic_formalism}: Presentation of the field equations in the context of EGB gravity. Section \ref{Hot_Casimir_Wormholes}: Construction of hot Casimir wormholes. Within this section, we have the following subsections: Subsection \ref{Geometric_Properties}: Discussion on embedding diagrams, providing a visual representation of the wormhole geometry. Analysis of the curvature scalar, highlighting the geometric properties of the solution. Subsection \ref{Physical_Conditions}: Analysis of the energy conditions and calculation of the amount of exotic matter, determining its role. In Subsection \ref{Stability_Complexity}: Investigation of the stability of the wormhole solutions. And stil, the calculation of the complexity factor, offering insights into the structural properties of the system. Finally, in Section \ref{Concluding_Remarks}: We present our final considerations and perspectives. Here, we adopt that $8\pi G_{D}=k_B=1$, where $G_{D}$ is the gravitational constant in $D$ dimensions, and the metric signature convention is $(-, +, +, +)$.


\section{Basic formalism of field Equations of EGB gravity}
\label{Basic_formalism}

In this section, we present the theoretical model of Einstein-Gauss-Bonnet gravity, which arises from the combination of the Einstein-Hilbert and Gauss-Bonnet terms in the gravitational action \cite{Lovelock, M.R.Mehdizadeh}. As mentioned earlier, these theories are of great interest since string theory predicts that, in the classical regime, Einstein's equations receive next-order corrections, typically represented by higher-order curvature terms in the action. Moreover, the Gauss-Bonnet term is the only quadratic curvature term that preserves the structure of second-order field equations.

Thus, in the context of EGB gravity, the gravitational action can be written as \cite{M.R.Mehdizadeh}:
\begin{equation} \label{gravity_action}
    \mathcal{I_{G}}=\int d^{D}x\sqrt{-g}\bigg(R+\mu_{2}\mathcal{G}\bigg),
\end{equation}
where $R$ is the Ricci scalar, $D$ represents the dimension of spacetime, and $\mu_{2}$ is the Gauss-Bonnet coefficient. The Gauss-Bonnet invariant, $\mathcal{G}$, is given by:
\begin{equation} \label{invariante_GB}
    \mathcal{G} = R^{2}-4R_{\alpha \beta}R^{\alpha \beta} + R_{\alpha \beta \delta \eta}R^{\alpha \beta \delta \eta} .
\end{equation}
Thus, by varying the action of Eq. \eqref{gravity_action} with respect to the metric tensor, the field equations are obtained by
\begin{equation} \label{field_equations}
    G_{\alpha \beta} + \mu_{2}\mathcal{H}_{\alpha \beta} = \mathcal{T}_{\alpha \beta},
\end{equation}
where $G_{\alpha \beta}$ is the usual Einstein tensor, $\mathcal{H}{\alpha \beta}$ is the Gauss-Bonnet tensor, and $\mathcal{T}{\alpha \beta}$ is the energy-momentum tensor. The expression for $\mathcal{H}_{\alpha \beta}$ is given by:
\begin{equation} \label{GB_tensor}
    \mathcal{H}_{\alpha \beta} = -\dfrac{1}{2}\mathcal{G}g_{\alpha \beta} + 2\bigg[RR_{\alpha \beta}-2R_{\alpha \sigma}R^{\sigma}_{\beta}-2R_{\alpha \rho \beta \nu}R^{\rho \nu} - 2R_{\alpha a b c }R^{a b c}_{\beta}\bigg].
\end{equation}

Now, we will analyze the static and spherically symmetric wormhole metric in $D-2$ dimensions, known as the Morris-Thorne metric \cite{M.S.Morris}, given by:
\begin{equation} \label{Morris_Thorne_metric}
    ds^{2} = -e^{2\Phi(r)}dt^{2} + \bigg(1-\dfrac{b(r)}{r}\bigg)^{-1}dr^{2} + r^{2}d\Omega^{2}_{D-2},
\end{equation}
where $\Omega^{2}_{D-2}$ denotes the spherical line element in $D-2$. The gravitational redshift function, denoted by 
$\Phi(r)$, describes the variation in the frequency of a photon as it escapes from a gravitational potential. Extracting a photon from this potential requires energy, which is proportional to its frequency, so $\Phi(r)$ represents the increase in frequency as the photon's energy grows. Therefore, if a wormhole has an event horizon, a photon may not have enough energy to escape. To prevent the formation of horizons, the redshift function must remain finite throughout its domain. The function $b(r)$ is the shape function, responsible for determining the geometry of the wormhole. The shape of the wormhole can be visualized through the embedding diagram \cite{M.S.Morris}. The radial coordinate $r$ is non-monotonic, varying in the range $r_0<r<\infty$, where $r_0$ represents the throat of the wormhole. The throat is defined as the region that connects the two ends of the wormhole. At the position $r_0$, the condition $b(r_0)=r_0$ must be satisfied. Moreover, we must consider two additional conditions: the flaring-out condition, which requires $b'(r_{0})<1$ or, equivalently,
\begin{equation} \label{flaring_out_condition}
    \dfrac{b(r)-rb'(r)}{b(r)^{2}} > 0,
\end{equation}
where the symbol $\prime$ denotes the derivative with respect to $r$, and the asymptotic flatness condition, given by
\begin{equation} \label{flatness_condition}
    \dfrac{b(r)}{r} \rightarrow 0 \ \ \ \text{as} \ \ \  r \rightarrow \infty.
\end{equation}

The energy-momentum tensor, given by Eq. \eqref{field_equations}, corresponding to an isotropic perfect fluid, is expressed as:
\begin{equation} \label{energy_tensor}
    T_{\alpha}^{\beta} = \text{diag}\left(-\rho(r), p_{r}(r), p_{t}(r), p_{t}(r), \cdots \right),
\end{equation}
where $\rho(r)$ is the energy density, $p_{r}(r)$ represents the radial pressure, and $p_{t}(r)$ are the transverse pressures. Thus, using Eqs. \eqref{field_equations} and \eqref{energy_tensor}, the field equations in EGB gravity can be expressed as:
\begin{eqnarray}
    \rho(r) &=& \dfrac{(D-2)}{2r^{2}}\bigg[\bigg(b'-\dfrac{b}{r}\bigg)\bigg(1+\dfrac{2\mu b}{r^{3}}\bigg)+\dfrac{b}{r}\bigg((D-3)+(D-5)\dfrac{\mu b}{r^{3}}\bigg)\bigg], \label{fields_equations_1} \\
    p_{r}(r) &=& \dfrac{(D-2)}{2r}\bigg[2\Phi'\bigg(1-\dfrac{b}{r}\bigg)\bigg(1+\dfrac{2\mu b}{r^{3}}\bigg)-\dfrac{b}{r^{2}}\bigg((D-3)+(D-5)\dfrac{\mu   b}{r^{3}}\bigg)\bigg], \label{fields_equations_2} \\
    p_{t}(r) &=& \bigg(1+\dfrac{2\mu b}{r^{3}}\bigg)\bigg(1-\dfrac{b}{r}\bigg)\bigg[\Phi''+\Phi'^{2}+\dfrac{(b-rb')\Phi'}{2r(r-b)}\bigg]+\bigg(\dfrac{b-b'r}{2r^{2}(r-b)}+\dfrac{\Phi'}{r}\bigg)\bigg(1-\dfrac{b}{r}\bigg) \nonumber \\ &\times& \bigg[(D-3)+(D-5)\dfrac{2\mu b}{r^{3}}\bigg]-\dfrac{b}{2r^{3}}\bigg[(D-3)(D-4)+(D-5)(D-6)\dfrac{\mu b}{r^{3}}\bigg] \label{fields_equations_3} \\ & - & \dfrac{2\Phi'\mu}{r^{4}}(D-5)(b-b'r)\bigg(1-\dfrac{b}{r}\bigg) \nonumber .
\end{eqnarray}
Here, for convenience, we define $\mu=(D-3)(D-4)\mu_{2}$. Additionally, we observe that the system consists of three field equations, given by Eqs. \eqref{fields_equations_1}, \eqref{fields_equations_2}, and \eqref{fields_equations_3}, and five unknowns: $\rho(r)$, $p_r(r)$, $p_t(r)$, $\Phi(r)$, and $b(r)$. This makes the system underdetermined, requiring additional techniques to obtain wormhole solutions. In the next section, we will analyze these functions within the framework of an approach based on Casimir energy. 


\section{Hot Casimir Wormholes in higher dimensions}
\label{Hot_Casimir_Wormholes}

One of the main areas of research in theoretical physics in recent times has been the problem of infinite vacuum energy in quantum fields \cite{Weinberg:1988cp}. This issue arises from calculations in Quantum Field Theory, which, even in the vacuum state where, {\it a priori}, there are no particles, predicts an unlimited amount of energy. Solving this dilemma is essential to understanding the nature of the vacuum and its impact on fundamental concepts such as the expansion of the universe and the amount of exotic matter in wormholes. To better understand and deal with these divergences, efforts have been made through various studies, including renormalization techniques and modifications to preexisting theories.

In the case of higher dimensions, the value of the Casimir energy density between two parallel plates is given by \cite{Wolfram, Svaiter, Milton}:
\begin{equation} \label{Casimir_energy_density}
    \rho(r) = -\dfrac{\lambda_{D}}{r^{D}} ,
\end{equation}
where the constant $\lambda_{D}$ is defined by
\begin{equation}
   \lambda_D = 2 \Gamma\left(\frac{D}{2}\right)(4\pi)^{-D/2}\zeta(D).
\end{equation}
From Eqs. \eqref{fields_equations_1} and \eqref{Casimir_energy_density}, adding the thermal correction term $-K_DT/r^{D-1}$, we obtain
\begin{equation} \label{fields_equations_subs}
    \frac{(D{-}2) \{ b(r) \left[ 2 \mu r b'(r){+}(D{-}4) r^3\right] {+}r^4 b'(r){+}(D{-}7) \mu b(r)^2\} }{2 r^6}+\frac{\lambda_D}{r^D}+\frac{k_D T}{r^{D-1}} = 0,
\end{equation}
where $k_D = 2 k_B (D-2)(2\sqrt{\pi})^{1-D} \Gamma[(D-1)/2] \zeta(D-1)$, with $k_B$ being the Boltzmann constant. Therefore, by solving Eq. \eqref{fields_equations_subs} for $b(r)$, we obtain:
\begin{equation} \label{shape_function_f}
    b(r) = -\frac{r^3}{2 \mu} \pm \frac{r^{3-D}}{2\mu} \sqrt{\frac{r^D[8\mu \lambda_D+(D-2)r^D+4 \mu^2(D-2)r f_1-8 \mu r k_D T \log{(r)}]}{D-2}},
\end{equation}
where $f_1$ is the integration constant. Applying the wormhole throat condition, $b(r_0)=r_0$, we obtain:
\begin{equation} \label{f1_constant}
    f_{1} = \frac{-2 r_{0}^5 T k_D \log (r_{0})+D r_{0}^{D+2}-2 r_{0}^{D+2}+D \mu  r_{0}^D-2 \mu  r_{0}^D+2 r_{0}^4 \lambda _D}{(D-2) \mu  r_{0}^5} .
\end{equation}
Therefore, the final structure of the shape function, $b(r)$, is obtained by substituting Eq. \eqref{f1_constant} into Eq. \eqref{shape_function_f}:
\begin{eqnarray} \label{shape_function} \nonumber
    b(r) &=& -\dfrac{r^{3-D}}{2 \mu}\bigg[r^{D} \pm r^{4}\bigg(\dfrac{r^{D}}{(D-2)r^{8}}\bigg(8 \mu  r T k_D \log (r)-8 \mu  \lambda_D+r^{D}(D-2)+\dfrac{4 \mu r}{r_{0}^{5}} \\ \label{shape_function_pn} & \times & -2 r_{0}^5 T k_D \log(r_{0})+D r_{0}^{D+2}-2 r_{0}^{D+2}+D \mu  r_{0}^D-2 \mu  r_{0}^D+2 r_{0}^4 \lambda_D\bigg)\bigg)^{1/2}\bigg].
\end{eqnarray}

At this point, it is important to note that Eq. \eqref{shape_function_pn} presents two possible shape functions: one with a positive multiplicative factor $(+)$ and another with a negative factor $(-)$. However, upon analyzing the asymptotic flatness condition, we find that only the shape function $b(r)$ with the negative factor satisfies this condition. Therefore, from this point onward, we will use the shape function with the negative sign for further analysis.

\begin{figure}[h!] 
\centering
\subfigure{\includegraphics[width=0.4\textwidth]{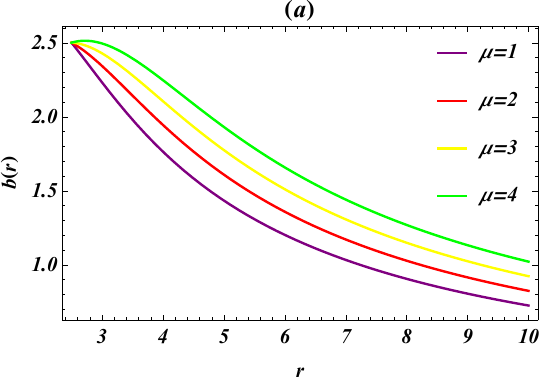}}
\subfigure{\includegraphics[width=0.4\textwidth]{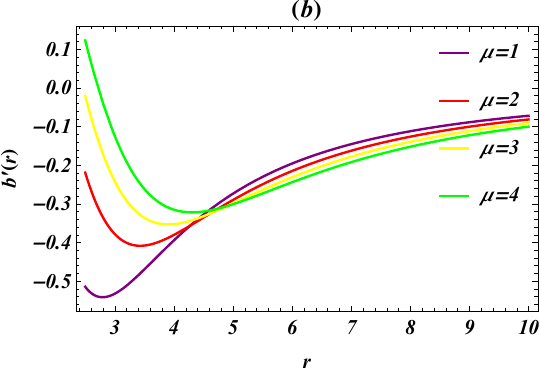}}
\caption{The plot $(a)$ shows dynamics of $b(r)$ versus $r$ whereas $b'(r)<1$ condition is depicted in plot $(b)$. Herein, we set $r_{0}=2.5$, $D=5$ and $T=1.2$}
\label{fig_1}
\end{figure}

\begin{figure}[h!]
\centering
\subfigure{\includegraphics[width=0.4\textwidth]{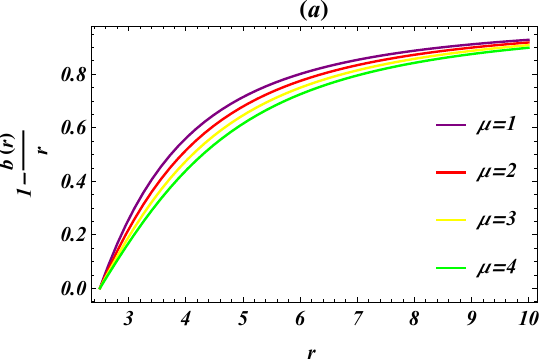}}
\subfigure{\includegraphics[width=0.4\textwidth]{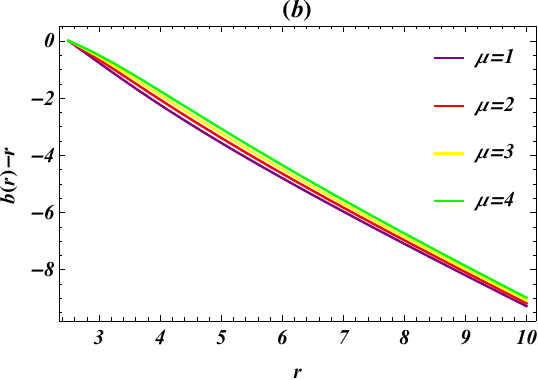}}
\caption{Plot $(a)$ shows dynamics of  $1-b(r)/r$ and plot $(b)$ displays dynamics of $b(r)-r$.}
\label{fig_2}
\end{figure}

\begin{figure}[h!]
\centering
\subfigure{\includegraphics[width=0.4\textwidth]{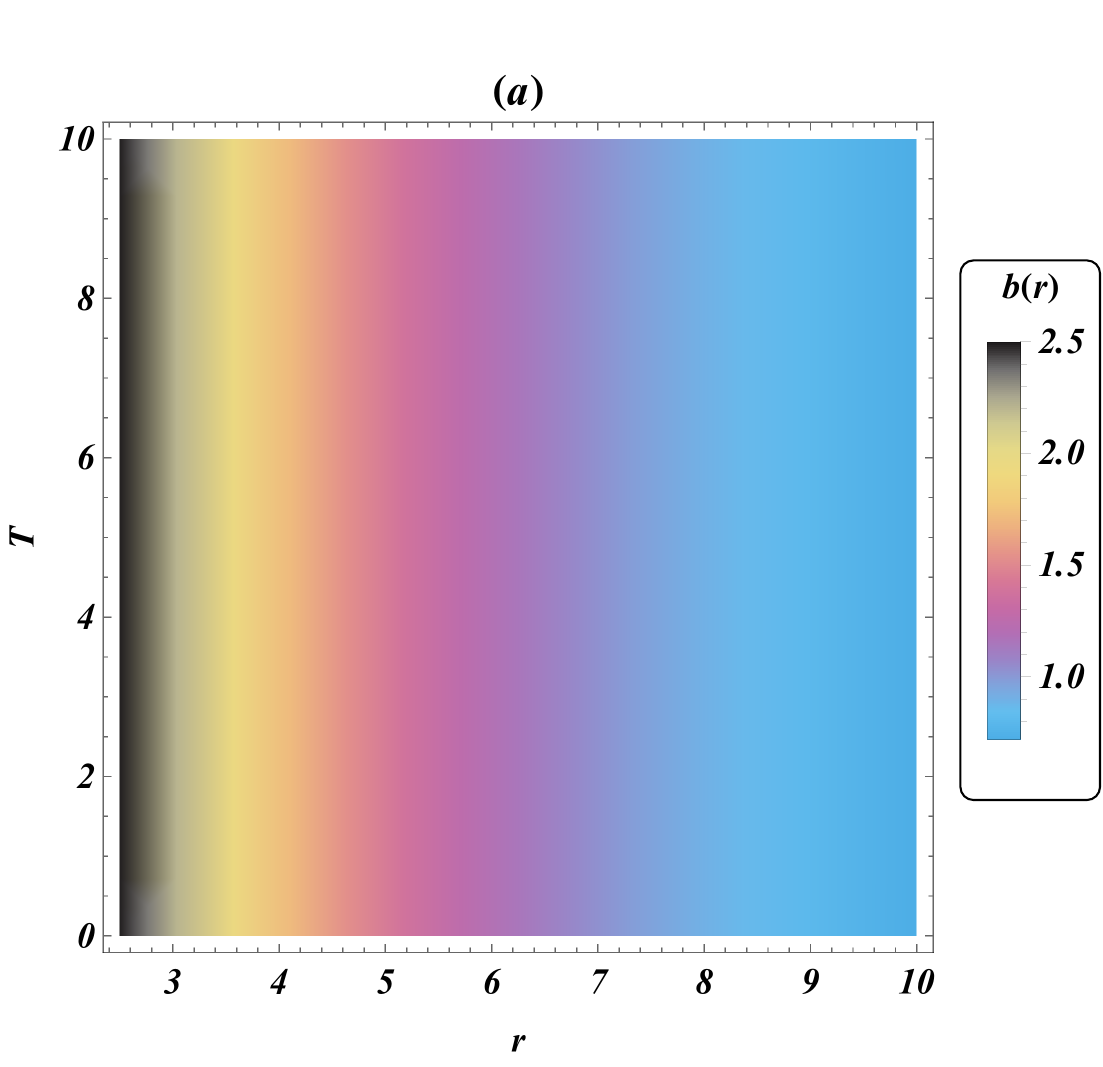}}
\subfigure{\includegraphics[width=0.4\textwidth]{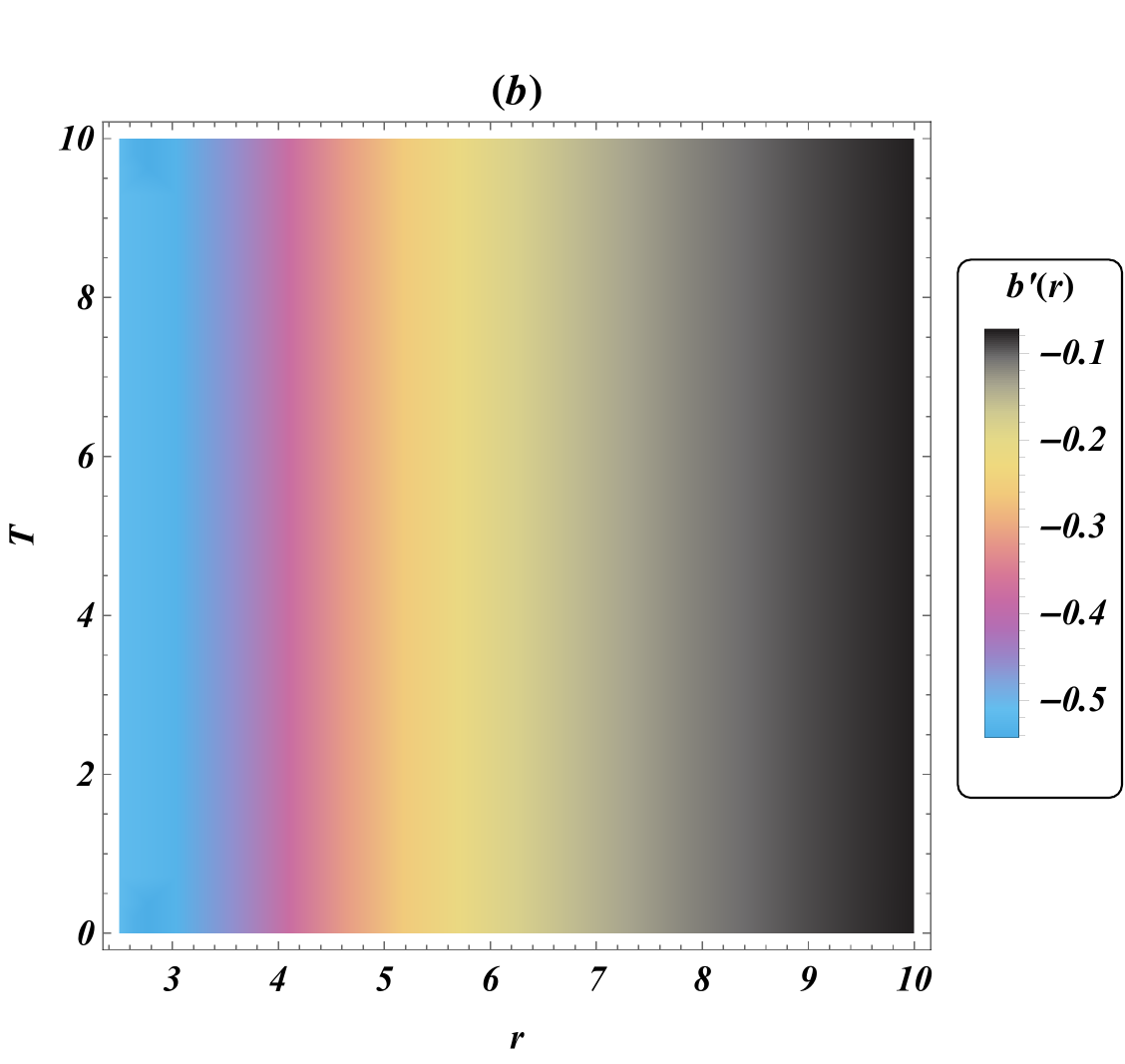}}
\caption{Plot $(a)$ illustrates the behavior of $b(r)$ as a function of $r$, while plot $(b)$ displays the condition $b'(r)<1$ using contour plots.}
\label{fig_3}
\end{figure}

\begin{figure}[h!]
\centering
\includegraphics[width=0.4\linewidth]{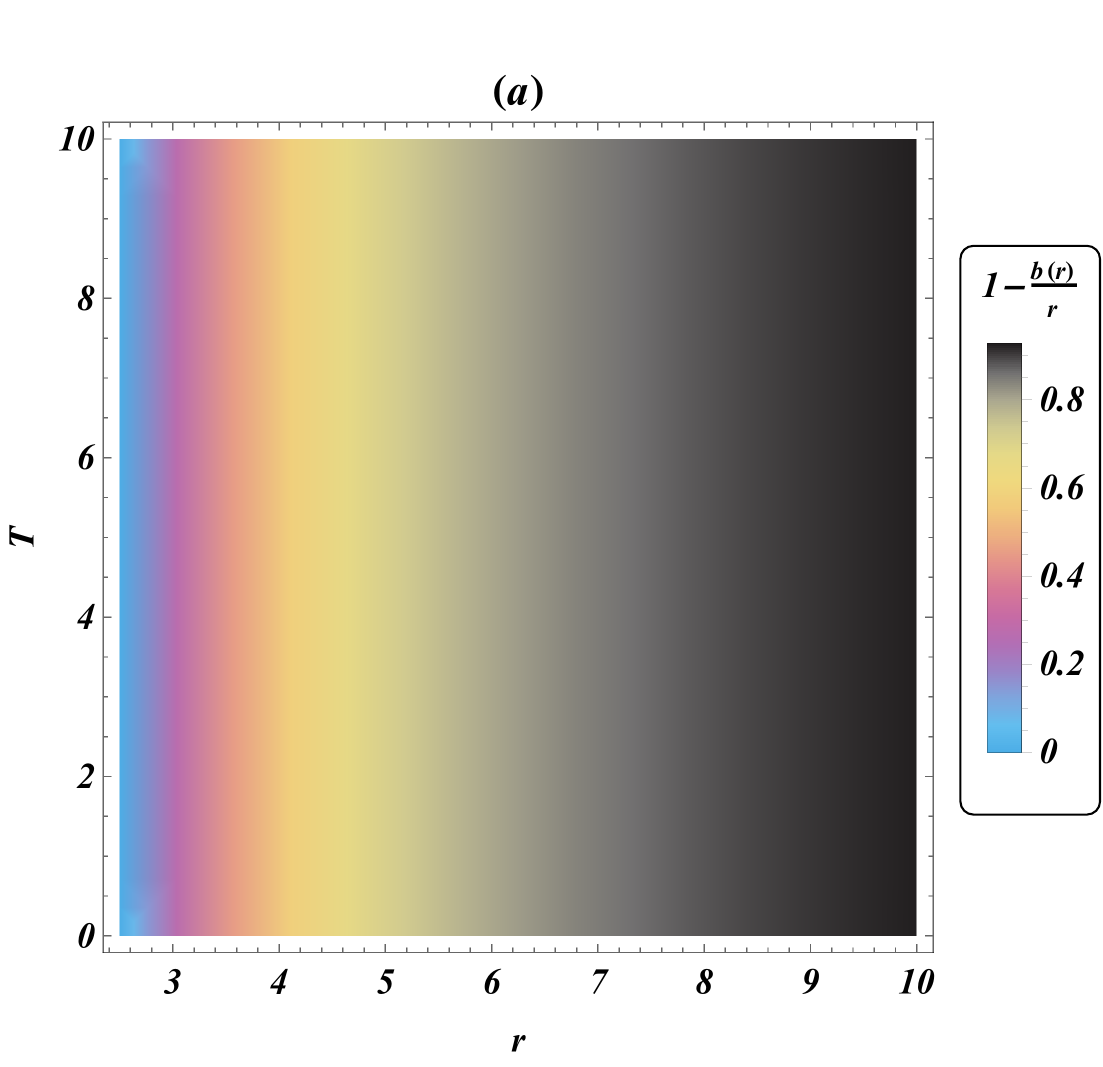}
\includegraphics[width=0.4\linewidth]{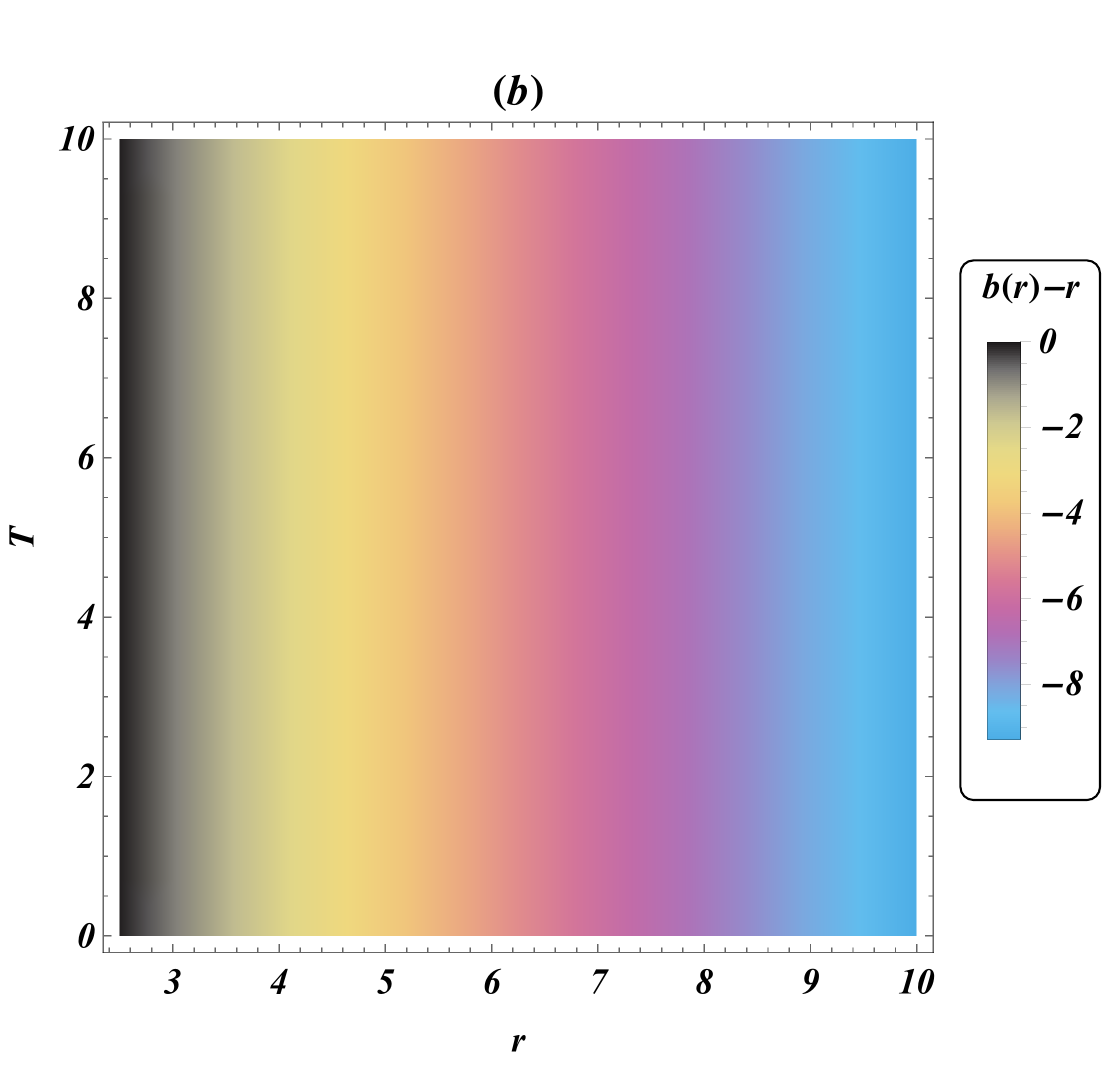}
\caption{Plot $(a)$ illustrates the dynamics of $1-b(r)/r$, while plot $(b)$ depicts the dynamics of $b(r)-r$.}
\label{fig_4}
\end{figure}

For a better visualization of the shape function, we present its behavior in Figs. \ref{fig_1}, \ref{fig_2}, \ref{fig_3}, and \ref{fig_4}. The Fig. \ref{fig_1} shows the dynamics of $b(r)$ and its derivative for different values of the parameter $\mu$, highlighting that the wormhole throat is located at $r=2.5$ and that the flaring-out condition is satisfied. In Fig. \ref{fig_2}, it is observed that the asymptotic flatness condition is clearly met. In Figs. \ref{fig_3} and \ref{fig_4}, we analyze the behavior of the shape function concerning the temperature $T$, where it is noted that near the throat, $b(r)$ assumes higher values.

\subsection{Geometric Properties}
\label{Geometric_Properties}
\subsubsection{Embedding diagrams}

\begin{figure}[h!]
    \centering
    \includegraphics[width=0.4\linewidth]{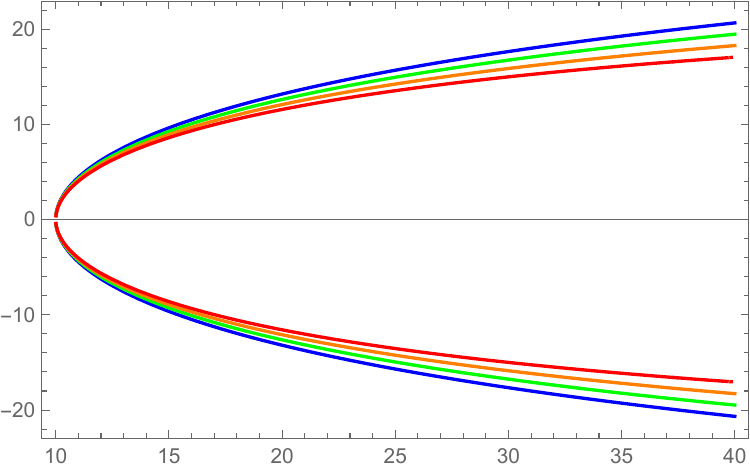}    \includegraphics[width=0.4\linewidth]{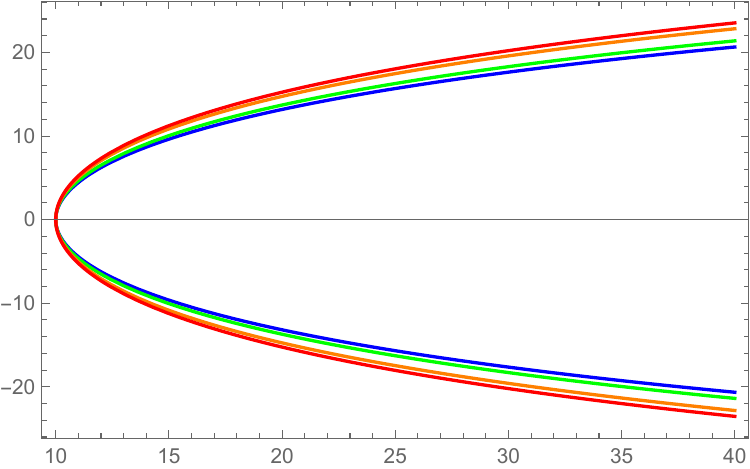}
    \caption{Left panel: Profile of embedding diagrams, for $D=5$, $\mu=0.2$, and $r_0=10$, considering the absolute temperatures $T=0$ (in blue), $T=500$ (green), $T=1000$ (orange), and $T=1500$ (red). Right panel: Profiles for $\mu=0.1$ (in blue), $\mu=5$ (green), $\mu=15$ (orange), 
 $\mu=20$ (red) considering $T=1000$, with the remaining parameters equal.}
    \label{fig01}
\end{figure}

In Fig. \ref{fig01}, we plot some profiles of embedding diagrams for the EGB-Casimir wormholes with temperature correction, which are generated from the mapping of the metric spatial sector in cylindrical coordinates at the equatorial plane, via
\begin{equation}
   dr^2+r^2d\phi^2+dz^2=\frac{dr^2}{1-\frac{b(r)}{r}}+r^2d\phi^2\Rightarrow z(r)=\int_{r_0}^{r}\left[\frac{b(u)/u}{1-b(u)/u}\right]^{1/2}du.
\end{equation}

The four curves in the left panel represent the wormhole's embedding diagrams at different temperatures: \(T = 0\) (blue), \(T = 500\) (green), \(T = 1000\) (orange), and \(T = 1500\) (red), corresponding to a 5-dimensional spacetime (\(D = 5\)), with an EGB parameter \(\mu = 0.2\) and a throat radius \(r_0 = 10\). As the temperature increases, the curvature near the throat becomes more pronounced, indicating that the thermal corrections steepen the wormhole's shape there, making it sharper at higher temperatures. This shows that the thermal correction of the Casimir energy significantly influences the wormhole's structure. In contrast, the right panel shows the effect of varying the EGB parameter \(\mu\) on the wormhole's geometry. As \(\mu\) decreases, the spatial curvature near the throat (\(r_0 = 10\)) becomes more pronounced, with steeper slopes for smaller values of \(\mu\). However, as the radial distance from the throat increases, the profiles flatten, suggesting a reduction in the spatial curvature far from the throat. Thus, the quantum parameter \(\mu\) smooths the curvature near the throat, while higher temperatures enhance it.

\subsubsection{Curvature scalar}

The Ricci scalar provides valuable insights into the local curvature of spacetime, helping to identify regions of high curvature or potential singularities. This analysis goes beyond the investigation of singularities based solely on the divergence in Eq. \eqref{shape_function}, offering a more comprehensive understanding of the wormhole's structure and stability. By studying the Ricci scalar, we gain a deeper understanding of the underlying physics, including the energy conditions and their violations, which are often associated with exotic matter necessary to sustain traversable wormholes. Thus, from Eqs. \eqref{Morris_Thorne_metric} and \eqref{shape_function}, it is possible to obtain:
\begin{eqnarray}
    \begin{aligned}
        R_5 = & \frac{r^{12} \left(\frac{1}{\mu r^8}\right)^{3/2}}{80 \pi r_0 \sqrt{r^5 \left(4 \pi^4 r T k_B (\log(r_0)-\log(r))+45 \left(\frac{8 \pi^2 r \left(r^4+4 \mu  \left(\mu +r_0^2\right)\right)}{\mu }-\frac{r \zeta(5)}{r_0}+\zeta(5)\right)\right)}} \\ & \times \Bigg \{ \sqrt{10} \left(45 \left(64 \pi^2 r r_0 \left(r^4+2 \mu \left(\mu +r_0^2\right)\right)+\mu \zeta (5) (3 r_0-4 r)\right)-4 \pi^4 \mu  r r_0 T k_B \right. \\ & \times \left. (4 \log(r)-4 \log(r_0)+1)\right) - 480 \pi \mu r^4 r_0 \sqrt{\frac{1}{\mu r^8}} \Bigg[r^5 \left(4 \pi^4 r T k_B (\log (r_0)-\log (r)) \right. \\ & \left. +45 \left(\frac{8 \pi^2 r
        \left(r^4+4 \mu \left(\mu+r_0^2\right)\right)}{\mu }-\frac{r \zeta(5)}{r_0}+\zeta(5)\right)\right)\Bigg]^{1/2} \Bigg \}.
   \end{aligned}
\end{eqnarray}

\begin{figure}[h!] 
    \centering
    \includegraphics[width=0.45\linewidth]{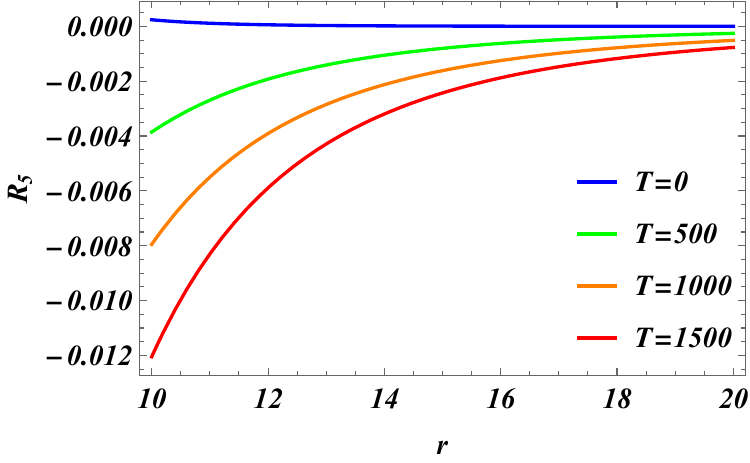}    \includegraphics[width=0.45\linewidth]{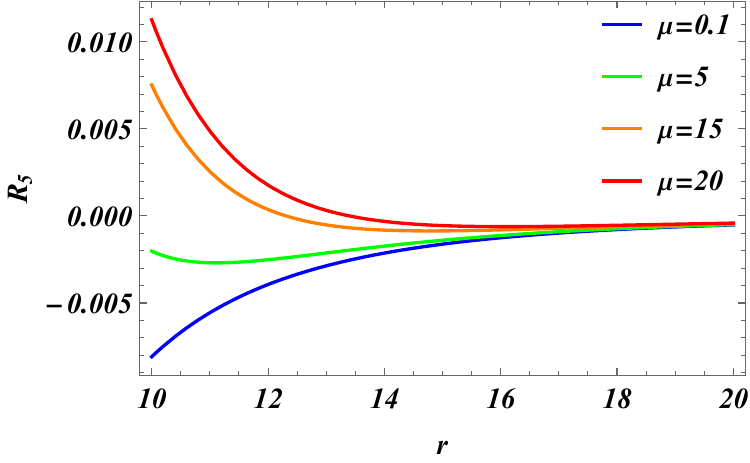}
    \caption{Left panel: Profile of Ricci scalar, for $D=5$, $\mu=0.2$, and $r_0=10$, considering the absolute temperatures $T=0$ (in blue), $T=500$ (green), $T=1000$ (orange), and $T=1500$ (red). Right panel: Profiles for $\mu=0.1$ (in blue), $\mu=5$ (green), $\mu=15$ (orange), $\mu=20$ (red) considering $T=1000$, with the remaining parameters equal.}
    \label{fig_ricci}
\end{figure}

Note that, in this case, we have specialized the calculation for $D=5$. For a more detailed understanding of the behavior of the Ricci scalar, we have graphically represented its variation in Figs. \ref{fig_ricci}. They emphasize that the solution is indeed asymptotically flat and free from divergences.

\subsection{Physical Conditions}
\label{Physical_Conditions}
\subsubsection{Energy conditions}

\begin{figure}[h!]
    \centering
    \includegraphics[width=0.4\linewidth]{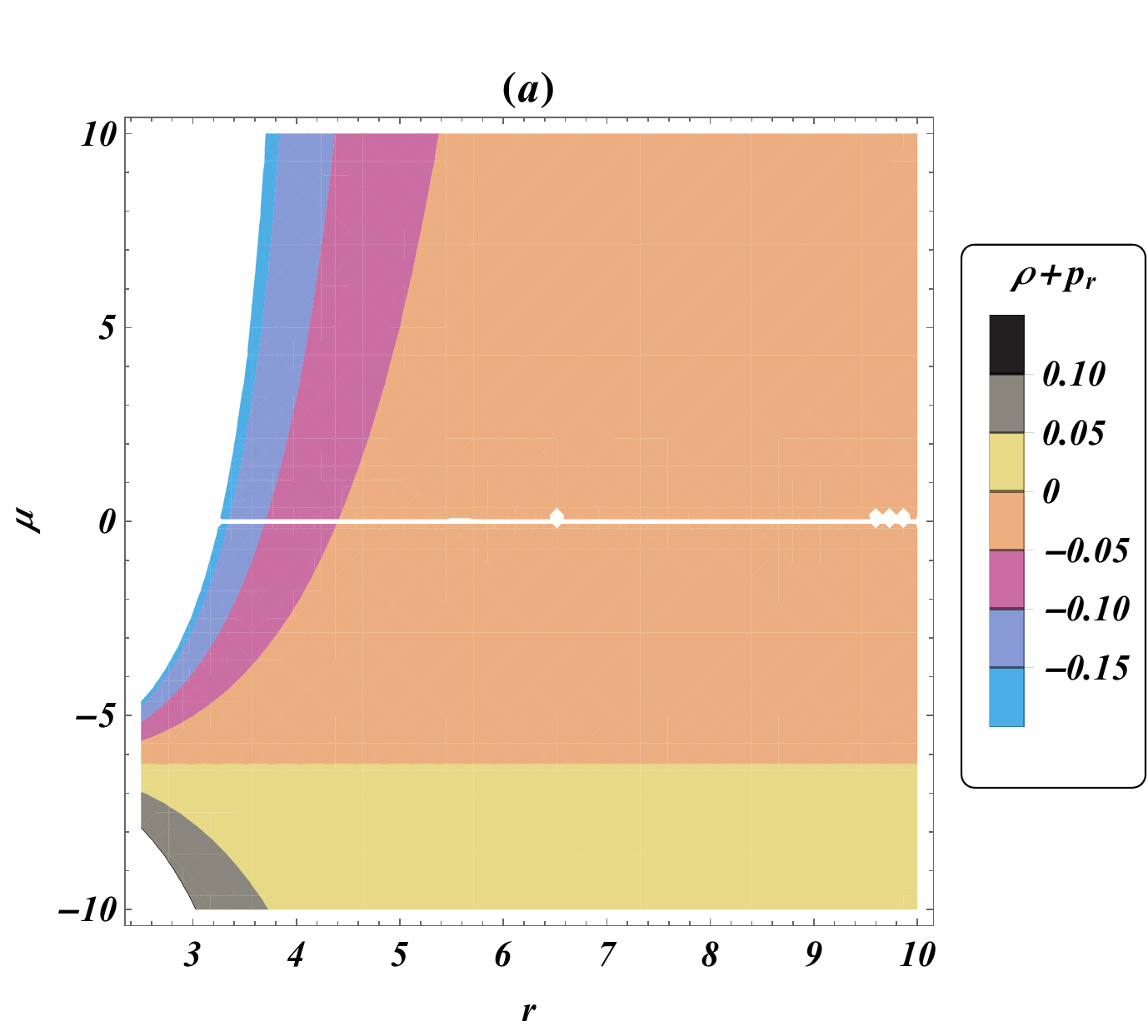}    \includegraphics[width=0.4\linewidth]{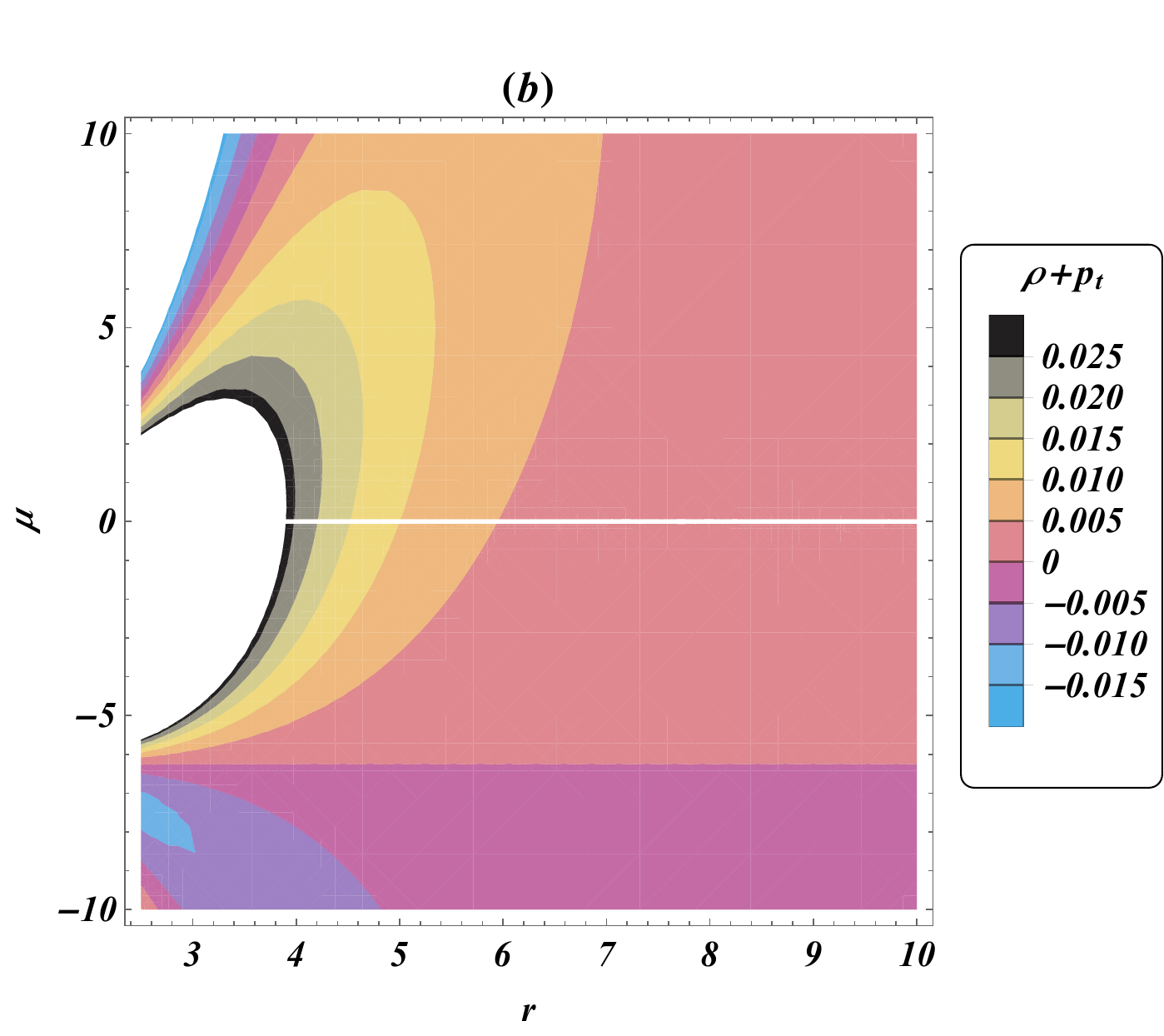}
    \caption{Dynamics of NEC in certain domains of $\mu$ and $T$ are fixed.}
    \label{fig.E1}
\end{figure}

\begin{figure}[h!]
    \centering
    \includegraphics[width=0.4\linewidth]{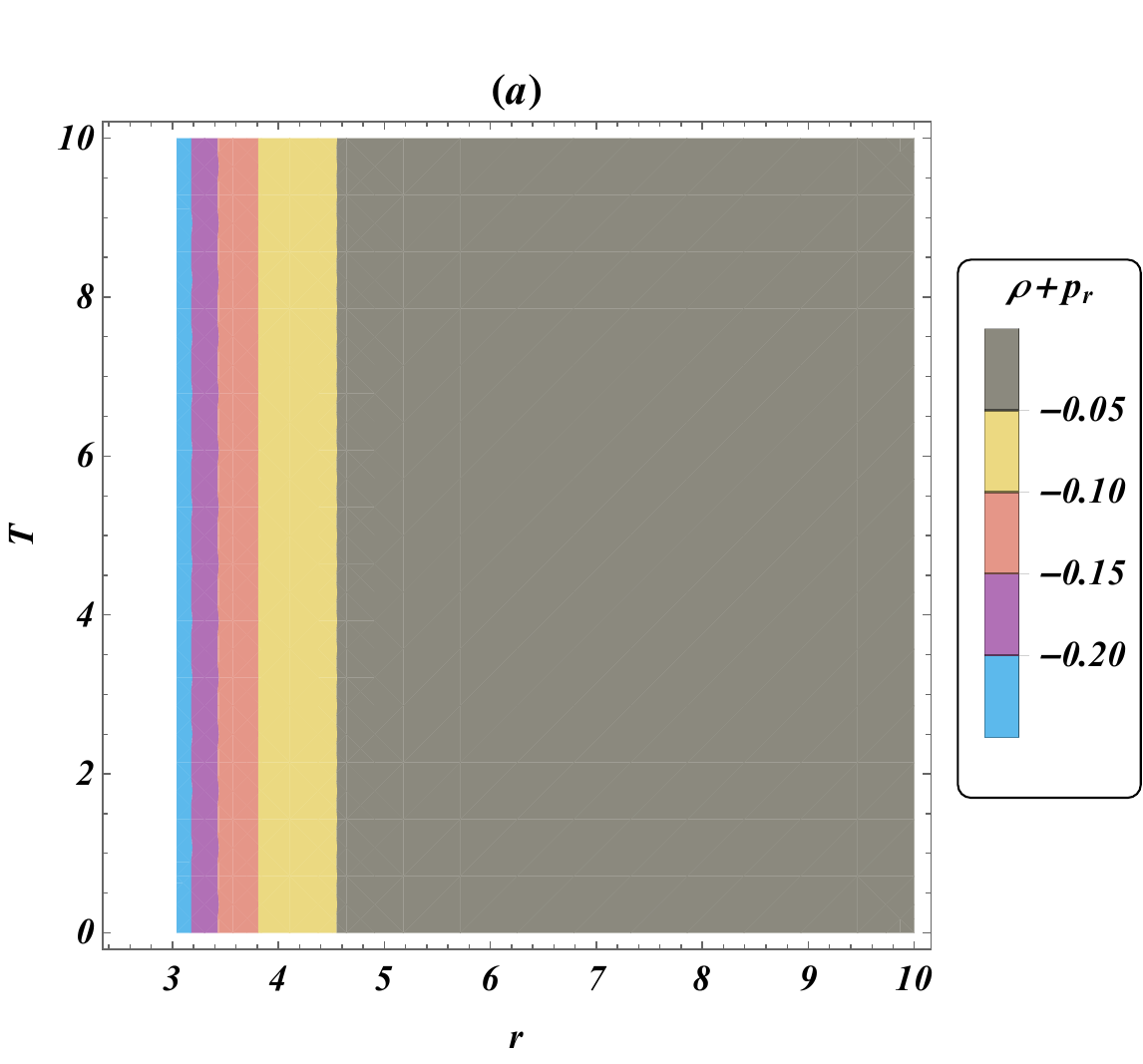}    \includegraphics[width=0.4\linewidth]{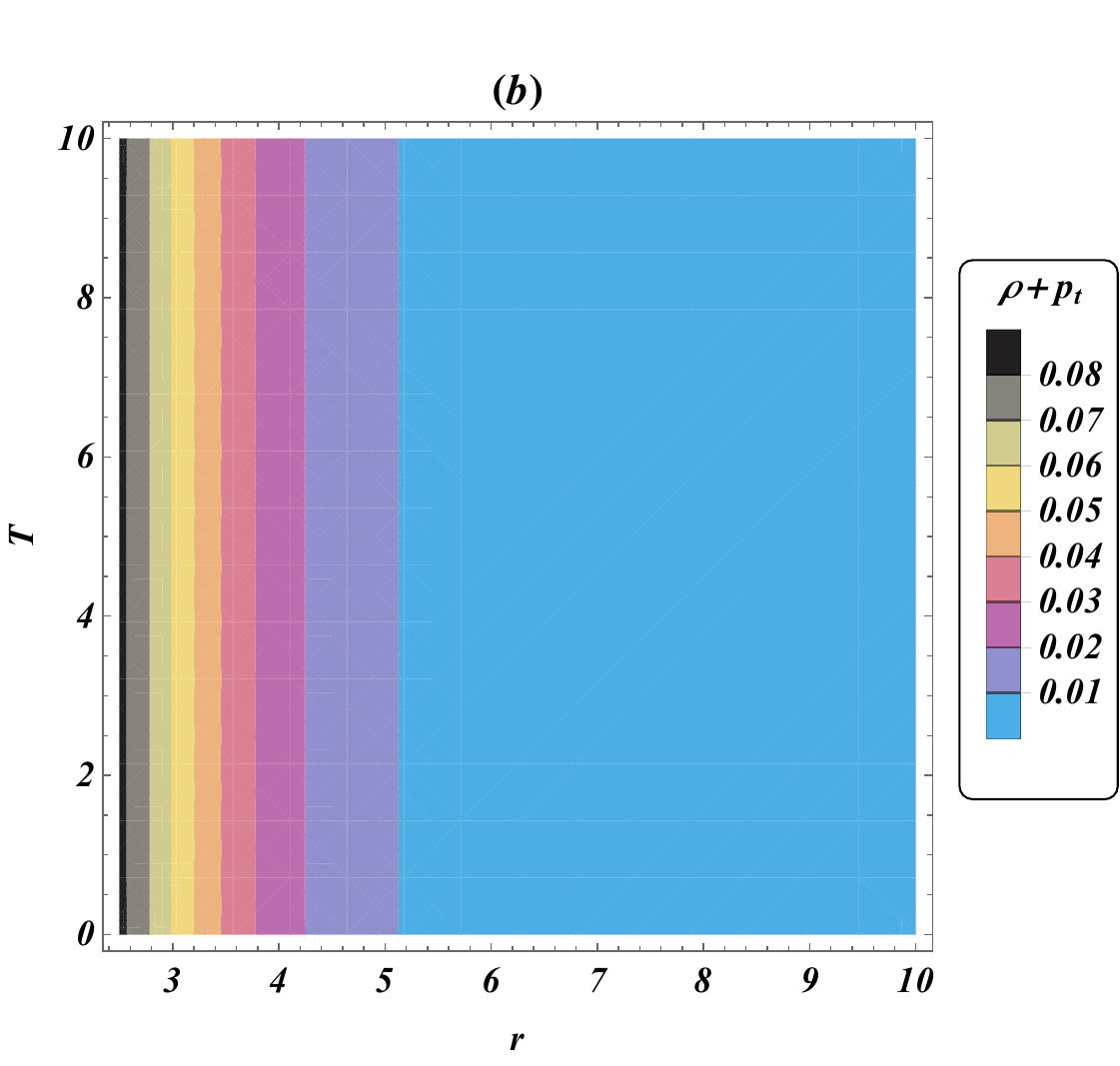}
    \caption{Dynamics of NEC in certain domains of $\mu$ and $T$ are fixed.}
    \label{fig.E2}
\end{figure}

It is a well known fact that, in GR, static traversable wormholes in 4D fails the energy conditions at or near the wormhole throat \cite{M.S.Morris,S. Kar}. The failure of energy conditions depends upon flaring out condition. The violation of energy conditions is closely linked with the flaring-out condition, which is a critical requirement for the structure of a traversable wormhole. Specifically, the flaring-out condition ensures that the throat of wormhole remains open, allowing for traversal, but this necessitates the existence of negative or exotic energy density at the throat. As a result, satisfying the flaring-out condition inherently leads to the failure of traditional energy conditions, such as the Null Energy Condition (NEC) and the Weak Energy Condition (WEC), in the vicinity of the wormhole throat. This connection highlights the necessity of exotic matter to sustain a stable and open wormhole geometry.

Next, we examined the energy conditions for the developed shape function, Eq. \eqref{shape_function}. To discuss these conditions the expressions for $\rho+p_{r}$ and $\rho+p_{t}$ are as follows:
\begin{eqnarray}
    \rho+p_{r} &=& \dfrac{2 \lambda _5 \mu +2 k_5 \mu  r T+3 r^5-\sqrt{3} \mu  r^4 \sqrt{\frac{1}{\mu  r^8}}  \mathcal{E}}{2 \mu  r^5}, \\
    \rho+p_{t} &=& \frac{k_5 r T+\lambda _5}{r^5}+\dfrac{r^{12} \left(\frac{1}{\mu  r^8}\right)^{3/2}}{6r_{0} \mathcal{E}}\times\bigg[-4 \sqrt{3} k_5 \mu  r r_{0} T (2 \log (r/r_{0})+1)+4 \sqrt{3} \lambda_5 \\ &\times& \mu (r_{0}-2 r) - 3r r_{0}\{3 \sqrt{3} r^4+4 \sqrt{3} \mu  \left(\mu +\text{r1}^2\right)-3 \mu  r^3 \sqrt{\frac{1}{\mu  r^8}} \mathcal{E}\}\bigg],
\end{eqnarray}
with 
\begin{eqnarray}
    \mathcal{E}=\sqrt{r^5 \left(8 k_5 r T (\log (r/r_{0}))+\frac{3 r \left(r^4+4 \mu  \left(\mu +r_{0}^2\right)\right)}{\mu }+\lambda _5 \left(\frac{8 r}{r_{0}}-8\right)\right)}.
\end{eqnarray}

The NEC are analyzed in two distinct ways. First, we plotted the behavior of the NEC over a specific range of the GB coupling parameter while keeping the temperature profile fixed, as shown in Fig. \ref{fig.E1}. We can see that $\rho+p_{r}>0$ for $\mu<-6$ and $\rho+p_{t}>0$ for $\mu>-6$, therefore we could not find the common region for $\rho+p_{r}$ and $\rho+p_{t}$ to be positive. Secondly, we investigated the behavior of the energy conditions over a certain range of temperature $T$ keeping the GB coupling parameter fixed as showsn in Fig. \ref{fig.E2}. We can observed from plot $(a)$ for all values of temperature $rho+p_{r}<0$, Hence, NEC fails.

The NEC is the most fundamental requirement among the various energy conditions, serving as the minimum criterion that must be met for energy conditions to hold. However, a violation of the NEC is a clear indication of the presence of exotic matter material with properties that differ significantly from conventional matter, such as negative energy density or negative pressure. This exotic matter is necessary for phenomena like traversable wormholes or other non-standard gravitational effects, where traditional notions of energy and causality are challenged.

\subsubsection{Quantity of exotic matter}

We are now going to analyze the Volume Integral Quantifier (VIQ), defined by \cite{Nandi:2004}:
\begin{equation} \label{viq}
    \mathcal{I}_v=\int_{r_0}^r \frac{4\pi^{(D-2)/2}}{\Gamma{[(D-1)/2]}} (\rho+p_r)x^{D-2}dx,
\end{equation}
calculated in $D$ spacetime dimensions, where $r$ represents an arbitrary radius near the throat, a region where NEC is commonly violated. The objective is to obtain the amount of exotic matter necessary to keep the wormhole throat open in the scenario under analysis. Thus, we have taken the integral on the energy density and radial pressure.

\begin{figure}[h!]
\centering
\subfigure{\includegraphics[width=0.48\textwidth]{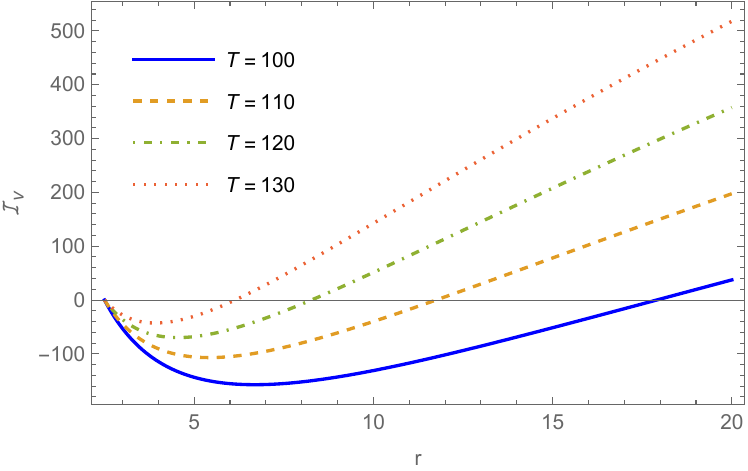}}
\subfigure{\includegraphics[width=0.48\textwidth]{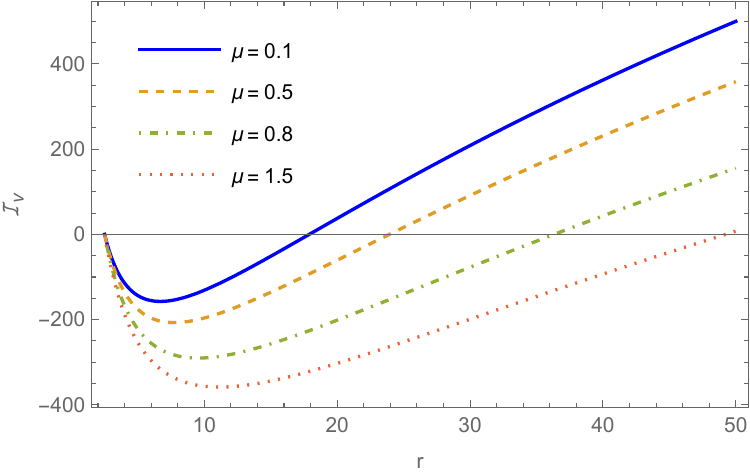}}
\caption{Radial dependence of VIQ, for varying parameters, with $r_0=2.5$ and $D=5$. Left panel: $\mathcal{I}_v$ for different temperatures \(T\), with other parameters fixed ($\mu=0.1$). Right panel: $\mathcal{I}_v$ for different values of the EGB \(\mu > 0\) parameter, considering $T=150$. 
 \label{quant}}
\end{figure}

The behavior of the volume integral quantifier (VIQ) as a function of the radial coordinate is shown in Fig. \ref{quant}. In the left panel, it is evident that as the wormhole becomes hotter (higher temperatures), VIQ decreases, indicating that less exotic matter is necessary to maintain the wormhole's geometry. This suggests that thermal energy partially compensates for the stabilizing effects typically attributed to exotic matter, reducing its overall requirement when other parameters are kept constant. In the right panel, we observe that increasing the EGB parameter $\mu$ leads to a larger (negative) VIQ. This result implies that higher-order curvature corrections, encoded by such a parameter, necessitate more exotic matter to counteract their influence and sustain the wormhole structure. These results reveal a complex interaction between temperature, curvature corrections, and exotic matter.

\subsection{Stability and Complexity}
\label{Stability_Complexity}
\subsubsection{Stability of the solutions}

\begin{figure}[h!]
\centering
\subfigure{\includegraphics[width=0.48\textwidth]{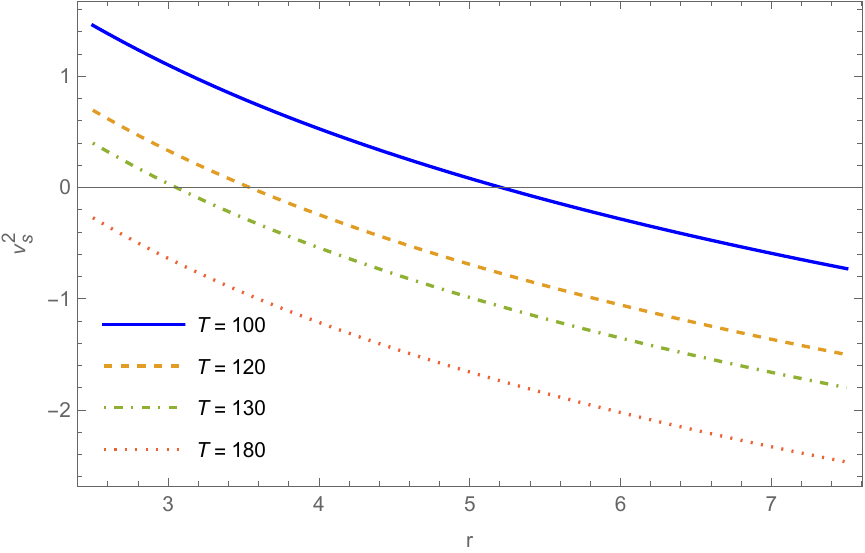}}
\subfigure{\includegraphics[width=0.48\textwidth]{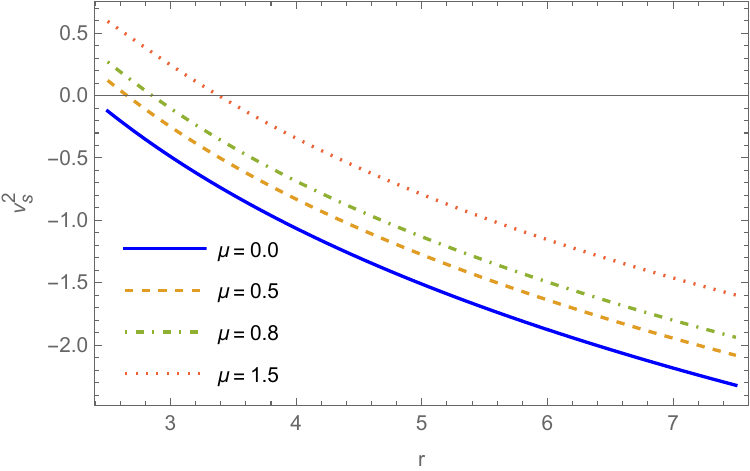}}
\caption{Radial dependence of the sound speed squared, \(v_s^2\), for varying parameters, with $r_0=2.5$ and $D=5$. Left panel: \(v_s^2\) as a function of \(r\) for different temperatures \(T\), with other parameters fixed ($\mu=0.1$). Right panel: \(v_s^2\) as a function of \(r\) for different values of the EGB \(\mu > 0\) parameter, considering $T=150$. 
\label{sound}}
\end{figure}

We will now investigate the stability of the obtained wormhole solution by analyzing the squared sound speed of the fluid along the radial direction \cite{Capozziello:2022zoz}. This quantity, denoted as $(v_s)^2$, can be calculated as follows:
\begin{equation}
  \label{vsound}
    (v_s)^2=\frac{d p_r}{d\rho_e}=\frac{d p_r/dr}{d\rho/dr}.
\end{equation}
The stability of the wormhole is assured if $(v_s)^2 > 0$, and it must satisfy $(v_s)^2 < 1$ to be physically meaningful. In the left panel of Fig. \ref{sound}, we illustrate the radial dependence of the radial sound speed squared, \(v_s^2\), while varying the absolute temperature \(T\), with all other parameters held constant. The results reveal a nuanced relationship between temperature and wormhole stability. At moderate temperatures, \(v_s^2\) remains positive and within physical bounds, $0 \leq v_s^2 \leq 1$, near the wormhole throat, indicating stability as the thermal energy effectively counterbalances gravitational forces. However, at very low temperatures, \(v_s^2\) exceeds the physical upper limit (\(v_s^2 > 1\)), resulting in non-physical sound speeds that violate causality. Conversely, at higher temperatures, \(v_s^2\) becomes negative near the throat, signifying instability as the thermal effects fail to sustain the wormhole structure. This analysis highlights the existence of an intermediate temperature range—a ``warm'' wormhole regime—where stability is maintained and the configuration remains physically viable.

We examine again \(v_s^2\) as a function of the radial coordinate \(r\), now varying the EGB parameter \(\mu > 0\). For larger values of \(\mu\), the squared sound speed \(v_s^2\) remains within physical limits near the wormhole throat, indicating that the EGB corrections enhance stability by amplifying curvature effects. Conversely, as \(\mu\) decreases, \(v_s^2\) becomes negative near the throat, signaling instability. This occurs because weaker EGB corrections diminish the effective curvature, disrupting the delicate balance between exotic matter and the geometric structure necessary to sustain the wormhole. In the limit of General Relativity (\(\mu = 0\)), this instability becomes particularly pronounced. The EGB parameter \(\mu\) thus governs the strength of higher-order curvature contributions: larger \(\mu\) intensifies curvature effects and stabilizes the wormhole, while smaller \(\mu\) leads to destabilization.

The temperature also plays a critical role: excessive thermal energy induces instability, whereas extremely low temperatures result in nonphysical configurations due to causality violations. These results highlight the existence of a ``stability window'' for both \(\mu\) and \(T\), within which the wormhole throat remains physically viable and dynamically stable.

\subsubsection{Factor of Complexity}

To quantify the internal structure and behavior of self-gravitating systems, Herrera \cite{L.Herrera} established the concept of complexity in relativistic astrophysical systems. The complexity factor was developed as a tool to measure how far a system deviates from simplicity, e.g. an idealized homogeneous or isotropic model. It imprisons the subtle mutual dependencies between various physical characteristics of the system, such as inhomogeneities in the energy density and anisotropies in the pressure. The calculated expression for complexity factor, $Y_{TF}$, of our system is as follows:
\begin{eqnarray} \nonumber
    Y_{TF} &=& \dfrac{1}{2r^{6}}\bigg(-b(r) \left(r \left(3 (D-3) r^2-2 (D-5) \mu  b'(r)\right)+\left(D^2-7 D+10\right) \text{$\mu $b}(r)\right) \\ \nonumber &+& (D-3) r^4 b'(r)+ \left(D^2-13 D+40\right) \mu  b(r)^2\bigg)-\dfrac{1}{2r^{3}}\int_{r_{0}}^{r}\frac{D-2}{2 r^4}\bigg(r^5 b''(r)+(D-6)\\ \label{C11} &\times& r^4 b'(r)+2 \mu  r^2 b'(r)^2+r b(r) \left(r \left(2 \mu  b''(r)-3 (D-4) r\right)+2 (D-12) \mu  b'(r)\right) \\ \nonumber &-& 6 (D-7) \mu  b(r)^2\bigg).
\end{eqnarray}

\begin{figure}[h!]
    \centering
    \includegraphics[width=0.5\linewidth]{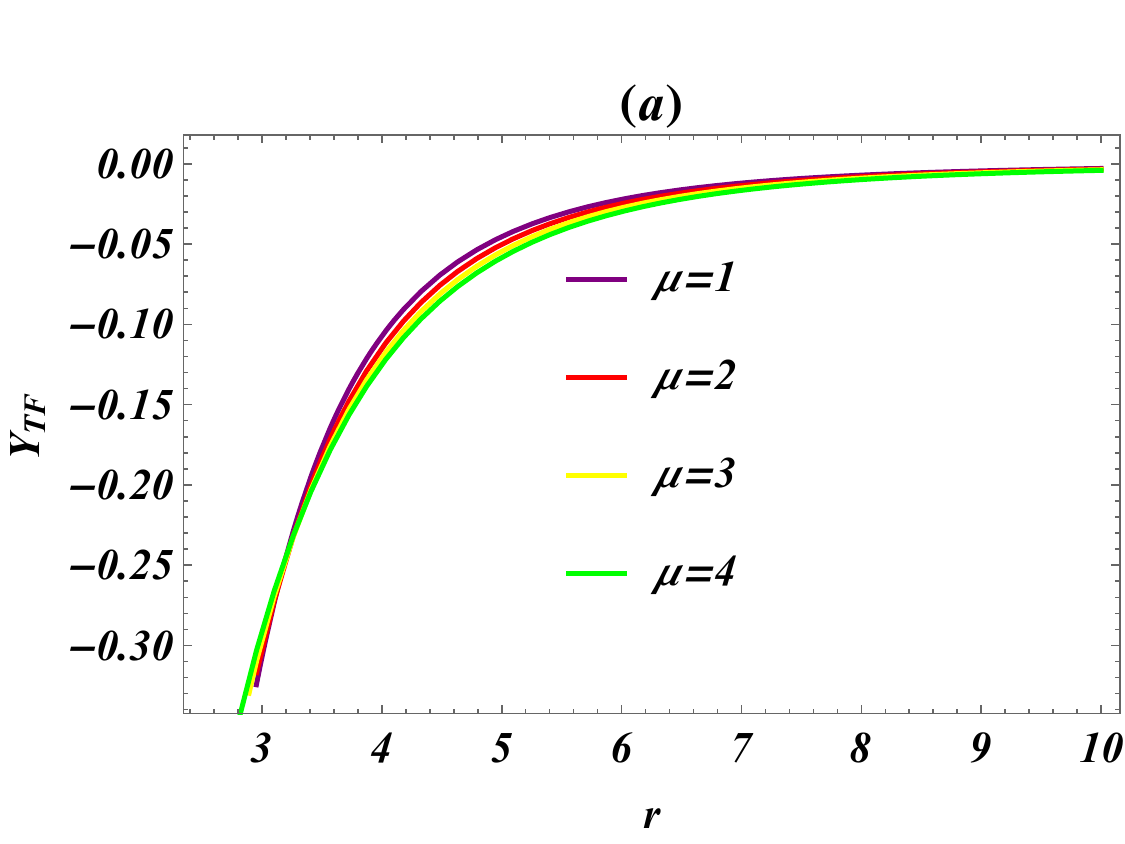}    \includegraphics[width=0.4\linewidth]{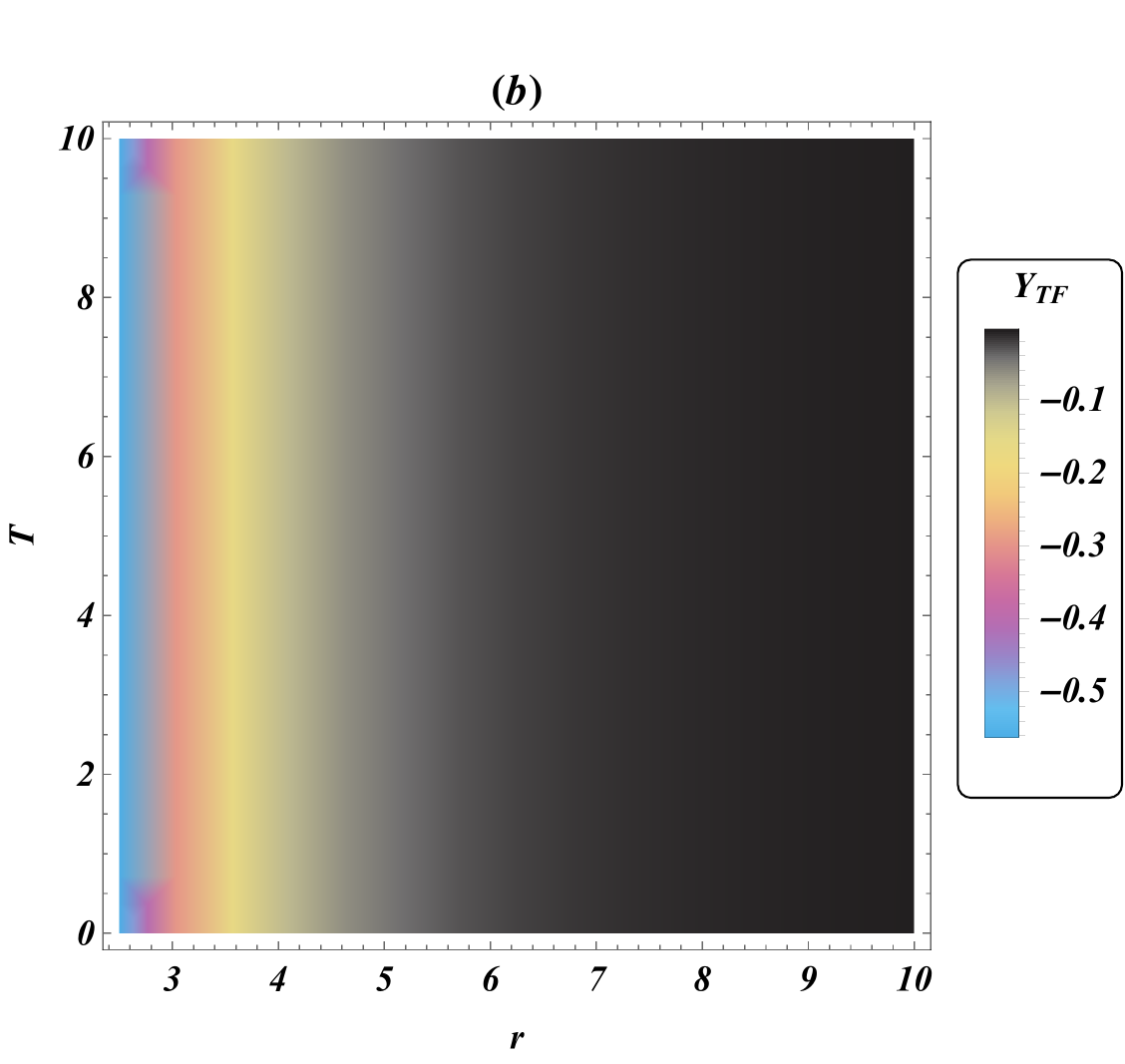}
    \caption{Plot $(a)$ illustrates the dynamics of the complexity factor with
    $T$ held constant, whereas plot $(b)$ demonstrates the behavior of the complexity factor with $\mu$ fixed. }
    \label{fig.C1}
\end{figure}

The behavior of complexity factor of casimir wormhole versus radial coordinate are shown in Fig. \ref{fig.C1}. By substituting the constructed shape function into Eq. \eqref{C11}, we observe that the EGB coupling parameter exerts a more pronounced influence than the temperature profile. We have observed that $r \rightarrow \infty$, or away from wormhole throat $Y_{TF} \rightarrow 0$. Still, the Fig. \ref{fig.C1} illustrates the impact of the temperature profile on the system's complexity. From this, we can assess that similar behavior is observed across all values of $T$ in regions near the wormhole throat, indicating a consistent complexity pattern close to this central area.

However, as the radial coordinate increases beyond a certain threshold, the complexity factor gradually diminishes, eventually reaching negligible values at larger distances from the throat. This suggests that the influence of temperature on complexity is most significant in close proximity to the throat and fades with radial distance. The minimal complexity factor exhibits homogenous energy density and isotropic pressure \cite{L.Herrera}. Furthermore, as long as these two effects cancel each other out on the complexity factor, the zero complexity factor predicts both anisotropic pressure and inhomogeneous energy density. Hence, in the vicinity wormhole throat, the complexity factor is monotonically increasing, and for higher values of radial coordinate, $Y_{TF}$ approaches zero. It has also been observed that for $\mu=10$, the energy density is homogenous at very high values of $r$, and pressure shows isotropic behaviour after $r>8$. Consequently, with greater values of the radial coordinate in the case of a wormhole, we observe that the complexity factor approaches 0. Furthermore, pressure isotropy is more important in the dynamics of the complexity factor than energy density homogeneity.


\section{Concluding Remarks}
\label{Concluding_Remarks}

In this article, we investigate the geometries of traversable, static, and asymptotically flat wormholes in the context of Einstein-Gauss-Bonnet gravity. The choice of this theory is motivated by its fundamental role in recent studies on a wide range of cosmological and astrophysical phenomena, including black holes, wormholes, quasi-normal modes, and thermal effects in higher-dimensional spacetimes.

Inspired by \cite{Remo Garattini2}, we consider the quantum vacuum energy density in the presence of two neutral conducting plates as the matter source for the wormhole. With this, we analyze the influence of thermal fluctuations on the Casimir effect and their impact on the global structure of the wormhole in EGB gravity. Our main objective is to understand how thermal contributions and quantum vacuum fluctuations interact to shape the wormhole geometry in this gravitational context.

From the EGB field equations and the thermally corrected Casimir energy density, we determine the shape function, Eq. \eqref{shape_function}. We verify that $b(r)$ satisfies all the necessary conditions for the existence of a wormhole, including asymptotic flatness, the throat conditions, and expansion requirements. Additionally, we perform a graphical analysis, Figs. \ref{fig_1}, \ref{fig_2}, \ref{fig_3}, \ref{fig_4}, \ref{fig01}, \ref{fig_ricci}, \ref{fig.E1}, \ref{fig.E2}, \ref{quant}, \ref{sound} and \ref{fig.C1}  to illustrate the variations of thermal fluctuations and investigate the internal dynamics of the wormhole in Einstein-Gauss-Bonnet gravity. All conditions associated with the shape function are examined through 2D plots, and we also present contour diagrams in a specific domain for a more detailed understanding of the influence of thermal fluctuations.

Our results indicate that the curvature near the wormhole’s throat increases with temperature, showing that thermal corrections enhance the wormhole's geometry. For small values of $\mu$, the spatial curvature near the throat becomes more pronounced, steepening as $\mu$ decreases. However, curvature profiles gradually become flatter as the radial distance from the throat increases, suggesting that spatial curvature diminishes with distance.

Overall, our findings suggest that hot Casimir wormholes can, in principle, remain traversable. In this context, it would be interesting to explore modified gravity corrections in higher dimensions, both for zero and finite temperatures. Furthermore, a promising approach would be to investigate the feasibility of new traversable wormhole solutions using the gravitational Casimir effect. Such a study could be conducted within both General Relativity and Einstein-Gauss-Bonnet gravity, allowing us to assess how the gravitational Casimir effect manifests in these theories.

Thus, this article contributes to the existing literature on Casimir wormholes in higher-dimensional spacetimes \cite{M. Zubair1}, advancing the study of quantum gravity effects on the stability of these structures and energy conditions. Additionally, within the framework of five-dimensional Einstein-Gauss-Bonnet gravity, the influence of noncommutativity on the existence of Casimir wormholes was analyzed in \cite{Mushayydha2}, demonstrating how this property of spacetime affects the shape function and curvature characteristics of the wormhole.

Finally, this work aims to correct inaccuracies present in \cite{M. Zubair1, Mushayydha, Mushayydha2}, where the expressions for Casimir wormholes were derived based on the four-dimensional formulation of the Casimir energy density. In contrast, we adopt here the appropriate formulation for Casimir wormholes in Einstein-Gauss-Bonnet gravity, considering the Casimir energy density in higher dimensions. Thus, this article has significant potential to contribute to the literature by providing a more rigorous approach and more precise results.


\section*{Acknowledgments}
\hspace{0.5cm} CRM thanks the Conselho Nacional de Desenvolvimento Cient\'{i}fico e Tecnol\'{o}gico (CNPq), Grants no. 308268/2021-6.


\end{document}